# Reconstruction of Air-Shower Parameters Through the Lateral Distribution Function of Ultra-High Energy Particles

Kadhom F. Fadhel[1,2, b], A. A. Al-Rubaiee[2, a]

[1] Directorate General of Education in Diyala, Ministry of Education, Baghdad, Iraq

[2] Department of Physics, College of Science, Mustansiriyah University, Baghdad, Iraq

[a] dr.rubaiee@uomustansiriyah.edu.iq, [b] kadhumfakhry@uomustansiriyah.edu.iq

**Abstract**

In this study, the necessity of the simulation study for exploring the interactions of ultra-high energy particles cosmic rays was examined. Different hadronic interaction models such as (SIBYLL, QGSJET, and EPOS) were simulated by using air showers simulation AIRES system (version 19.04.00). Also, the charged particle density of Extensive Air Showers (EAS) was calculated by estimating the lateral distribution function (LDF). Moreover, the LDF simulation of the two primary particles (proton and iron nuclei) was performed, taking into account their primary energies effect and the zenith angle of charged particles that produced in the EAS, within the energy range ($10^{17}$-$10^{19}$) eV. At extremely high energies ($10^{17}$, $10^{18}$, and $10^{19}$) eV, new parameters as a function of the primary energy were obtained by fitting the lateral distribution curves of EAS using Sigmoidal function (Logistic model). Comparison of the results showed a good agreement between the values obtained from the parameterized LDF using Sigmoidal function with experimental results by AGASA EAS observatory for the primaries proton as well iron nuclei, with the production of (electron positron) pair and the charged muons secondary particles at high energy about $10^{19}$ eV and ($\theta = 0^0$).

## 1. Introduction

The investigation of the characteristics of EAS triggered by ultra-high energy cosmic rays is crucial [1]. In addition, high-energy cosmic rays (CRs) have been detected through the chain reaction EAS that are produced in the air surrounding the Earth. Since certain primary particles are directly undetectable. So, they must be investigated based on the showers that measured in different ways [2]. It is observed, when any high-energy astronomical particle collides with an atom in the atmosphere surrounding the Earth, it generates a shower of secondary particles, which interact and generate more secondary particles before reaching the Earth's surface [3]. Primary CR properties must be deduced the particle and ratios in the shower and the creation of the shower in the atmosphere [2]. The study of the cascades that result from their interactions with atmospheric nuclei would provide an unique look at hadronic interaction properties at center-of-mass energies about one order of magnitude higher than those attained in human-made colliders [3]. A simple analytical model can't thoroughly explain the comprehensive shower creation because it is too complicated [4]. The reaction often called a cascade continues until the average energy possessed by these single and multiple particles drops below their critical energy, which is often lost due to multiple collisions rather than other radiative processes [5]. Therefore, it's normally modeled using a Monte Carlo (MC) simulation of each individual shower particle's transport and interaction, based on our current understanding of interactions, decays, and particle transport in matter [6]. Because of the complexities of the systems involved occur during the construction of an air shower, numerical simulations are often used to conduct comprehensive studies of its characteristics. So, affects both of the simulation of particle interactions and transport in the atmosphere, as well as the model assumptions, affect quantitative results [7]. All of the processes that have a major impact on the shower's actions must be taken into account by the simulating algorithms. So are all of electrodynamic interactions, hadronic collisions, photonuclear processes, particle decays, and so on [8, 9]. The density of charged particles as a function of the primary energies is seen in the current calculations. While, the simulation of the LDF performed using the AIRES system for (proton as well iron nuclei) primary particles with the energy range ($10^{17}$-$10^{19}$) eV for two different zenith angles ($0^0$ and $10^0$). The LDF of secondary particles at the production of electron-positron pair and the muons charged secondary particles was simulated. By the sigmoidal function (Logistic model) are obtained new parameters of LDF, the dependence of the densities for the particles produced in EAS as a function of the primary energy inside the energy spectrum range ($10^{17}$ - $10^{19}$) eV. The comparison gave a good agreement between the estimated LDF of the charged particles and AGASA EAS observatory as well the simulated results by Sciutto at $10^{19}$ eV.

## 2. The Lateral Distribution Function

The charged particle local density at different distances from the central position is one of the basic EAS characteristics that can be calculated very accurately by large ground-based air shower arrays. Since the detection of the EAS, the lateral or radial density distributions **ρ** (r) of various types of particles generated in EAS have been focused for experimental and theoretical researches [10, 11]. The right LDF of EAS muons and electrons is crucial for EAS research and study. The problem of quick LDF calculations that are both correct and adequate to the experimental data at very large distances from the shower axis (r ≥1 km), has yet to be solved. Therefore, the analysis of experimental data on giant air showers and the design of new experiments, reliable results over such large distances are needed [12].

The most widely used method for super-high energy EAS simulation gives an empirical overview of electromagnetic sub-showers based on various modifications. Therefore, the reconstruction of the shower core and shower path, knowledge of the LDF is essential. It can also be compared to model calculations for primary mass details and give useful information. The EAS lateral distributions are important for the air shower phenomenon for a variety of reasons. The first is that the energy and mass of the primary particle can be calculated and is the most important from the number and distribution of ground particles. To connect the observables to primary energy and mass, more accurate algorithms need comprehensive air shower simulations [13, 14].

The LDF is the shower characteristic of the cascade at different depths in the Earth's atmosphere [15]. The Nishimura-Kamata-Greisen (NKG) function, which is expressed via the following formula, is a commonly used term to describe the LDF type [16]:

$$\rho(R) = \frac{N_e}{2\pi R_M^2} C(s) \left(\frac{R}{R_M}\right)^{(s-2)} \left(\frac{R}{R_M} + 1\right)^{(s-4.5)} \qquad (1)$$

where $\rho_e(R)$ is the electrons density on the distance $r$ from the shower core; $N_e$ is the total number of electrons in the shower; $R_M$ is the Molier radius at sea level ($R_M$=78 m); $s$ is the shower age parameter where the NKG-function is valid for the range $0.8 < s < 1.6$, and $C(s)$ is the normalizing factor which is equal $0.366\ s^2\ (2.07 - s)^{1.25}$ [17].

## 3. Results and Discussion

### 3.1. AIRES Simulations

The specifics of shower evolution are extremely complicated to be completely delineated by uncomplicated analytical modeling. In addition, the MC simulation of interaction and transport of every single particle is needed to execute for accurate shower modeling evolution. Lately, MC packages are employed for simulating EAS using AIRES (AIR shower Extended Simulation) system [18]. Various hadronic interaction models are utilized for these event generators, such as SIBYLL [19], QGSJET [20] and EPOS [21]. Therefore, the air shower simulation programs are made up of a variety of interconnected procedures that run on a data set with a variable number of records, changing the contests and increasing or decreasing the size of the data set according to predetermined laws. Internal control procedures in AIRES' simulation engine continuously check and report particles touching the ground and/or moving over predetermined observing surfaces between the ground and injection stages. Where the number of showers is determined and then the identity of the elementary particle is determined, as well as its energy that will interact with the atoms of the atmosphere. Then we define the name of the task, as well as the kinetic energy of electrons, muons, and gamma rays. Next, we define the thinning energy and the zenith angle, and then choose the observing levels for the array to be used. And finally, we define the name of the secondary particles resulting from the chain reaction. The diffractive interactions possess a straight influence onto the shower progress. Also, that fact is clearly confirmed by graphing the densities of showers versus the shower core of atmosphere, at certain value of energies of $10^{17}$, $10^{18}$, and $10^{19}$ eV. The graphs were plotted depending on the data incoming from simulations executed via the AIRES system for assorted hadronic interaction models (SIBYLL, QGSJET and EPOS). The simulation was employed to investigate the production of primary particles (proton as well the iron nuclei) resulted from air showers within the range of primary energy ($10^{17}$-$10^{19}$) eV and explore the LDF growth of various hadronic interaction models created subsequent primary CRs of the extremely high value of energy react with the atmosphere and organize overall correlated production data [4]. Whole simulation process is executed via the AIRES system which benefited from the use of the thinning level $10^{-6}$ relative.

In figure 1 was displayed the difference between the primaries iron nuclei as well proton LDF at the primary energy $10^{19}$ eV and vertical EAS showers for two secondary particles, the production of the electron-positron pair and the muons charged secondary particles.

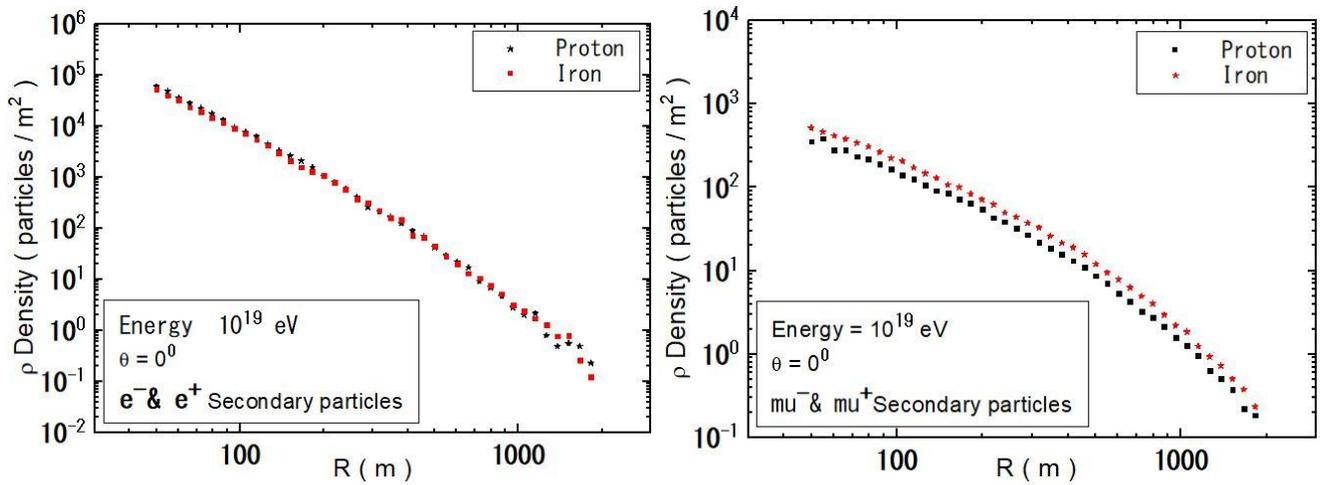

**Figure 1.** The simulation of the LDF by AIRES system for the primary proton as well iron nuclei at vertical showers.

In figure 2 was shown the simulation of LDF using AIRES system for different hadronic interaction models (SIBYLL, QGSJET, and EPOS) for the primaries (proton as well iron nuclei) at the fixed energy $10^{18}$ eV and inclined zenith angle ($\theta = 10°$), for muons charged secondary particles.

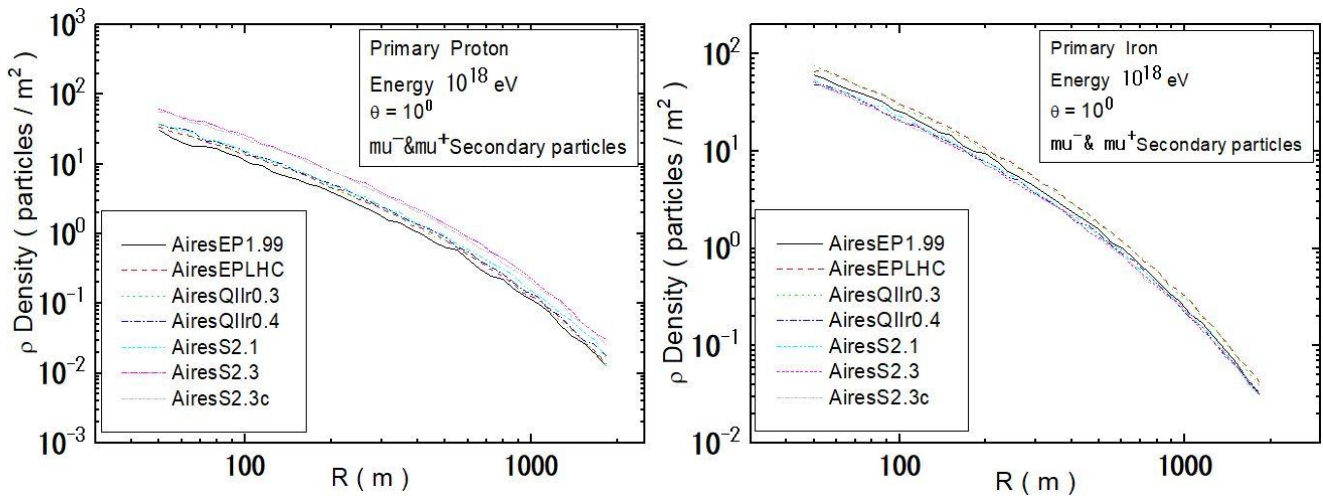

**Figure 2.** Comparison with different hadronic interaction models like (SIBYLL, QGSJET, and EPOS) by using AIRES system of LDF at the primary energy $10^{18}$ eV for secondary muons.

## 3.2. Parameterization of LDF

The Aires simulation was applied for the primary proton as well iron nuclei within the energy range ($10^{17}$-$10^{19}$) eV and explores the LDF for different hadronic models. A sigmoidal function (Logistic model) was used to parameterize the LDF of showers that started in EAS, yielding four new parameters for various primary particles, this function is denoted by:

$$\rho(E) = \frac{\eta - \zeta_c}{1+(\frac{x}{\delta})^\alpha} + \zeta_{c'} \tag{2}$$

when $\rho$ is the density of EAS showers as a function of primary energy; $\eta$, $\zeta_c$, $\delta$ and $\alpha$ are obtained coefficients for the LDF (see the Table 1). These coefficients are obtained by fitting the AIRES results, which are given by the polynomial form:

$$K(E) = a_o + a_1(E/eV) + a_2(E/eV)^2 \tag{3}$$

where $K(E) = \eta, \zeta_c, \delta, \alpha$ are parameters of Eq. (2) as a function of primaries and $a_o$, $a_1$ and $a_2$ are their coefficients (see the Table 1).

**Table 1.** The coefficients of the sigmoidal function (Logistic model) (Eq. 2) used to parameterize the AIRES simulation for the primaries proton as well iron nuclei within the energy range ($10^{17}$-$10^{19}$) eV and two zenith angles ($\theta = 0°$ and $10°$).

| Primary particles | Secondary particles | $K(E)$ eV | Coefficients | | |
|---|---|---|---|---|---|
| | | | $a_o$ | $a_1$ | $a_2$ |
| p | $e^-$ & $e^+$ | $\eta$ | 118.86495 | 1849.60556 | 26485.14832 |
| | | $\zeta_c$ | -0.9945 | -27.25218 | -142.30949 |
| | | $\delta$ | 86.23676 | 59.78296 | 46.00644 |
| | | $\alpha$ | 1.33741 | 1.21566 | 1.34987 |
| | mu$^-$ & mu$^+$ | $\eta$ | 2.26595 | 21.77746 | 153.23286 |
| | | $\zeta_c$ | 0.82059 | 8.07847 | 40.86455 |
| | | $\delta$ | 1253.03081 | 1094.68689 | 1101.96244 |
| | | $\alpha$ | 8.18762 | 7.46504 | 6.00806 |
| Fe | $e^-$ & $e^+$ | $\eta$ | 114.98125 | 1468.11536 | 23410.17673 |
| | | $\zeta_c$ | -0.86126 | -24.80306 | -208.61533 |
| | | $\delta$ | 83.17814 | 80.39021 | 49.32528 |
| | | $\alpha$ | 1.26734 | 1.28711 | 1.2592 |
| | mu$^-$ & mu$^+$ | $\eta$ | 3.29002 | 26.11111 | 217.12133 |
| | | $\zeta_c$ | 1.20671 | 9.54119 | 68.42865 |
| | | $\delta$ | 1169.71658 | 1102.07871 | 1102.75785 |
| | | $\alpha$ | 7.08781 | 6.27649 | 6.43017 |

Figure 3 shows the parameterization of the shower density in EAS as a function of primary energy using sigmoidal function (Logistic model) (Eq. 2) for two primary energies ($10^{17}$ and $10^{19}$) eV and two zenith angles ($0^0$ and $10^0$) at the production of the electron-positron pair and the muons charged secondary particles.

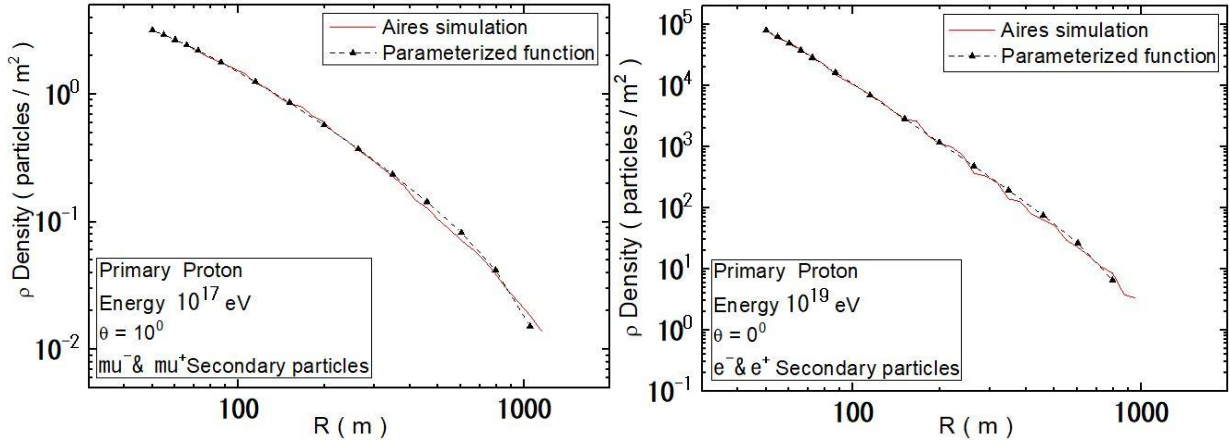

**Figure 3.** Lateral distribution that simulated with AIRES system (solid lines) and one calculated with Eq. (2) (scattered) for primary proton at energies $10^{17}$ and $10^{19}$ eV within the production of the electron-positron pair and the muons charged secondary particles.

### 3.3. The comparison with AGASA Observatory

The parameterized LDF that was obtained using Eq. 2 (sigmoidal function) was compared with the experimental results for the AGASA Array. This comparison gave a good agreement for both primary proton and iron nuclei at the fixed primary energy ($10^{19}$) eV for vertical EAS showers that initiated muons charged secondary particles [22].

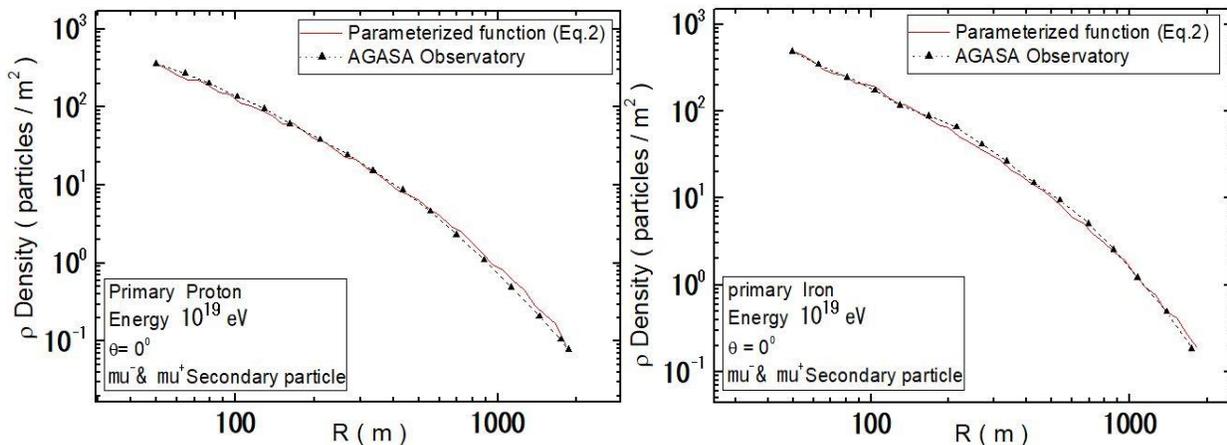

**Figure 4.** Comparison of the parameterized LDF obtained using (sigmoidal function) with the experimental results by AGASA Array for the primary proton and iron nuclei at the energy $10^{19}$ eV.

## Conclusions

The simulation of EAS lateral distribution function was performed by using AIRES system for three hadronic interaction models (SIBYLL, QGSJET, and EPOS) for primary proton and iron nuclei. The simulation was performed for three different high energies ($10^{17}$, $10^{18}$, and $10^{19}$) eV and two zenith angles ($0^0$ and $10^0$) of several secondary particles. In addition, it was calculated the parameters of LDF as a function of primary energy using the results of this simulation, through the sigmoidal function (logistic model) for the primary proton as well as iron nuclei at the energy range ($10^{17}$ - $10^{19}$) eV. The comparison of the parameterized lateral distribution function with that measured with the AGASA observatory demonstrates the ability to classify and calculate the energies of primary particles and determining around the ankle region of CR energy spectrum. The ability to build a library of lateral distribution samples that could be used for analyzing specific events observed with the EAS array and reconstruction of the primary CRs energy spectrum as well mass composition is the key benefit of the current method.

## Acknowledgments
Authors thank Mustansiriyah University in Baghdad- Iraq and AIRES system creators for their support in this work.